%% file: PD2025-1112-1.tex
\documentclass[12pt]{article}
\usepackage[utf8]{inputenc}
\usepackage[T1]{fontenc}
\DeclareUnicodeCharacter{202F}{\,}
\usepackage[margin=1in]{geometry}
\usepackage{amsmath,amssymb}
\usepackage{graphicx}
\usepackage{booktabs}
\usepackage{tabularx}
\usepackage{natbib}
\usepackage{url}  
\usepackage{tikz}
\usetikzlibrary{decorations.pathreplacing}
\graphicspath{{fig/}}
\begin{document} \title{The Effect of Punishment and Reward on Cooperation in a Prisoner's Dilemma Game} \author{Alexander Kangas} \date{November 12, 2025} \maketitle

\begin{abstract}
This paper characterizes how different incentive instruments shape cooperation in a repeated Prisoner`s Dilemma with a continuum of players.  A simple tit-for-tat strategy competes against unconditional defection, and the long-run outcome is summarized by a tipping-point share of cooperators,  above which cooperation spreads and below which defection prevails. Closed-form expressions for this tipping-point are derived as a function of four payoff classes: targeted punishment of defectors, general punishment applied to all deviations, and two symmetric reward instruments. The formula implies sharply diminishing returns to targeted punishment so that increasing the penalty lowers the temptation payoff and reduces the cooperation threshold but can only drive the threshold asymptotically toward zero, so a positive mass of defectors persists even under extreme sanctions.  By contrast, sufficiently strong general incentives (taxes, subsidies, or reputational payoffs) that shift the entire payoff profile can increase cooperation and make it self-enforcing. The framework nests several standard strategy refinements (evil and generous tit-for-tat, Win-Stay-Lose-Shift, heterogeneous horizons, and imperfect monitoring) and derives the corresponding shifts in the cooperation threshold. A numerical calibration illustrates these comparative statics and links them to policy environments where unilateral incentives conflict with collective welfare, including climate agreements, online platform governance, and systemic financial regulation.  Across these applications, the results suggest that durable cooperation is rarely secured by a single heavyweight mechanism, but rather that robust outcomes emerge when policy simultaneously tilts payoffs away from unilateral defection and extends the effective horizon over which cooperative gains are realized, combining moderate targeted measures with broader, general instruments.
\end{abstract}

\section{Introduction}
Why do individuals pay taxes, recycle when it is inconvenient, vote in low-stakes elections, or volunteer code to open-source projects when unilateral free-riding is often less personally costly? Such puzzles are central to research on social dilemmas across economics, political science, psychology, and evolutionary biology \citep{Ostrom1990, BowlesGintis2011, Nowak2011}. The two-player Prisoner`s Dilemma (PD) provides a stark representation of this puzzle where defection strictly dominates cooperation, yet mutual defection reduces joint welfare and a socially suboptimal outcome emerges from individually rational choices.

This study examines how punishment and reward shape cooperation in indefinitely repeated PDs. Motivated by insights from computational tournaments \citep{Axelrod1984} and evidence on costly punishment \citep{FehrGaechter2000, AmbrusGreiner2012}, a formal model is developed in which agents are randomly matched to play an indefinitely repeated game. Incentive mechanisms -- targeted or general, punitive or rewarding -- alter payoffs to strategic choices. \emph{Targeted} punishment or reward modifies a single payoff entry; \emph{general} punishment or reward applies a uniform tax or subsidy to all payoffs associated with a strategy type.

Relative to existing work, the contribution here is to place these incentive mechanisms and several common strategy extensions within a single, unified comparative-static framework that delivers closed-form thresholds for cooperation and transparent derivatives with respect to incentive intensities. The analysis distinguishes targeted from general instruments in a way that makes their different marginal effects explicit and comparably interpretable, shows how diminishing returns can arise for targeted punishment within the model, and clarifies conditions under which sufficiently strong general penalties or subsidies can fully deter defection in principle. The same lens is applied to forgiving and retaliatory strategies (e.g., Tit-for-Tat variants and Win–Stay, Lose–Shift), imperfect monitoring, and heterogeneous horizons, yielding parallel expressions that move the cooperation threshold in intuitive directions. Throughout, expressions are stated in general payoff terms, and any illustrative calibrations are clearly separated from the analytical results.

\subsection{Prior Literature}
Research on social dilemmas investigates how individually rational defection can be curbed such that cooperative outcomes emerge.  Foundational research highlighted the collective-action problem in one-shot or finite-horizon settings, where external enforcement is often necessary \citep{Olson1965, Hardin1968}.  Subsequent research demonstrated that repeated interaction, reciprocity, and strategic punishment can sustain cooperation \citep{AxelrodHamilton1981, Axelrod1984, Ostrom1990, BowlesGintis2011, FehrGaechter2000}.  Laboratory evidence further confirmed that both costly punishment and reward influence behavior in public-goods and prisoner's-dilemma environments \citep{Andreonietal2003, BallietMulderVanLange2011}.

From this foundation,  a first strand of literature compares the efficacy of sanctions and rewards.  Meta-analyses suggest that peer punishment is effective but subject to over-use and retaliation,  whereas rewards can indeed encourage cooperation \citep{BallietMulderVanLange2011} but they are susceptible to second-order free-riding \citep{Randetal2009}.  Field evidence indicates that indirect or reputational sanctions can amplify cooperative norms without the full monetary cost of direct fines \citep{BalafoutasNikiforakisRockenbach2014}.  From an evolutionary game-theoretic perspective, \citet{Yanetal2024} show that when incentive costs are borne by an external institution,  pure punishment collapses the defection basin by driving the cooperation threshold to zero, whereas pure rewards merely shrink it,  providing evidence that broad,  cost-free sanctions are more potent than rewards in sustaining cooperation.

A second strand addresses how the \textit{scope} of an incentive -- targeted versus general -- affects outcomes.  Targeted interventions that adjust a payoff entry deter exploitation in particular cases but display diminishing marginal returns once behavior adjusts (\cite{FehrGaechter2000}).  Broader,  strategy-wide incentives mirror formal institutions such as carbon pricing or blanket reputation scores,  which can,  in principle,  eliminate defection when sufficiently strong but impose information or implementation costs \citep{VanLangeRockenbachYamagishi2015}. 

A third strand of literature incorporates heterogeneous strategy sets and dynamic environments.  Laboratory evidence on generous tit--for--tat, forgiving win--stay--lose--shift,  and moral framing (e.g.,\ \cite{RandOhtsuki2013,NowakSigmund1993,CapraroRand2018}) shows that modest tweaks to classical strategies can significantly shift cooperation thresholds.  Complementary work on imperfect monitoring and private signals highlights the role of information accuracy in sustaining cooperation (\cite{ChendeVriesLensink2015}).

The present model attempts to unify these strands by placing incentives within a single comparative-static framework,  offering a theoretical basis for comparison within a unified analytical structure.  The paper concludes by mapping the comparative static results onto real-world levers, highlighting among other insights why moderate general incentives can outperform extreme targeted ones.  To preview these policy insights,  the model suggests that uniform carbon prices can often outperform case-by-case emission fines in curbing free-riding,  consistent with findings by \cite{Nordhaus2015}.  Similarly,  public health initiatives often incorporate all four incentive types,  providing a rich policy space to interpret the comparative static results.  Other illustrations include how platform-wide governance rules can stabilize cooperation more reliably than sporadic user bans in the technology space,  and why economy-wide counter-cyclical buffers might curb financial risks more effectively than targeted stress tests.  In each domain the model pinpoints which lever shifts the cooperation threshold most efficiently and how auxiliary design choices such as the level of forgiveness, monitoring accuracy, or horizon length amplify or dampen that shift.

\section{Model}\label{sec:model}
Consider an indefinitely repeated prisoner's dilemma game played by a finite population of $n$ players (with $n$ assumed large and,  for simplicity,  even).  In each period the population is randomly and anonymously partitioned into $n/2$ pairs,  each of which plays a prisoner's dilemma game.  Although the setting involves an $n$-player population,  each period ultimately features a two-player repeated interaction between the matched partners.  Because players do not know their partners' histories from previous matches,  every new match begins as a fresh encounter.

A standard one-shot prisoner's dilemma allows each player either to cooperate or to defect.  Mutual cooperation yields a payoff $R$ for each player, mutual defection yields a lower payoff $P$, and one-sided defection yields the highest payoff $T$ to the defector and the lowest payoff $S$ to the exploited cooperator.  The notation $\sigma = T$, $\gamma = R$, $\psi = P$, and $\omega = S$ is adopted, with $\sigma$ referred to as the \emph{temptation} payoff and $\omega$ as the \emph{sucker's} payoff.  The payoff ordering is therefore $\sigma > \gamma > \psi > \omega$.  All payoffs are assumed to be non-negative, and the additional condition $2\gamma > \sigma + \omega$ is imposed\footnote{Following \cite{Rapoport1965}.} so that the total payoff from mutual cooperation exceeds that from unilateral defection, ensuring the social efficiency of cooperation despite individual incentives to defect. 

For reference, the paper retains the $\{\sigma,\gamma,\psi,\omega\}$ notation because the comparative statics emphasize the temptation and sucker payoffs.  Readers who prefer the conventional $(T,R,P,S)$ symbols can transparently translate between the systems via
\[
T\equiv\sigma,\qquad
R\equiv\gamma,\qquad
P\equiv\psi,\qquad
S\equiv\omega,
\]
and the analysis below preserves the standard ordering $T>R>P>S$.  

\textbf{\noindent\emph{Notation note.}}  Throughout the policy sections the term ``punishment'' refers to the incentive levers $\theta$ or $\alpha$, whereas $\psi$ is reserved exclusively for the mutual-defection payoff ($P$ in the canonical PD notation).  This separation avoids conflating policy instruments with stage-game outcomes.

In the repeated version of the game,  each match between two players proceeds through a sequence of rounds.  In every round,  both players simultaneously choose whether to cooperate or defect and receive the corresponding stage-game payoffs.  The interaction terminates at the end of each round with probability $\rho$ (and continues with probability $1-\rho$), implying an expected match length of $1/\rho$ rounds.  For analytical simplicity,  no discounting beyond the continuation probability is introduced; future rounds within a match are treated as equally valued. 

When a match ends,  the two players are randomly re-matched with new partners,  and all prior history is discarded.  Each member of the population commits \textit{ex ante} to one of two strategies: a \textit{nice tit-for-tat} strategy ($T$) or an \textit{always-defect} strategy ($D$).\footnote{The `always-cooperate' strategy is omitted,  as any always-cooperator would earn a lower payoff than tit-for-tat in the presence of defectors and is thus dominated by TFT.}  A tit-for-tat player cooperates in the first round of any match and thereafter mimics the opponent's previous action, while an always-defect player defects in every round.  Because a $D$ type never cooperates first, a single round is sufficient to reveal the opponent's strategy: observing defection in the initial round signals a $D$ type, whereas observing cooperation weakly indicates $T$.  Once a match terminates and new partners are assigned, this information is reset, and each new encounter again starts under uncertainty about the other`s type. 

Incentive mechanisms are introduced next.  \emph{Targeted punishment} is defined as a sanction applied to defection in a specific circumstance.  Let $\theta\ge 0$ denote the intensity of a penalty levied on a defector who exploits a cooperator.  If a player defects while the opponent cooperates, the defector's payoff for that round is reduced by the fraction $\theta$, yielding $(1-\theta)\sigma$.  At $\theta=0$ no penalty is applied, whereas $\theta=1$ represents a sanction that drives the defector's opportunistic payoff to zero; values $\theta>1$ capture confiscatory fines that over-correct the temptation payoff and leave the defector with a negative net return, thereby breaking the canonical ordering $\sigma>\gamma>\psi>\omega$ that underpins the analysis.  Likewise, $\theta<0$ would raise the temptation payoff above $\sigma$ and effectively act as a targeted reward for defection.  

Perfect monitoring of opportunistic behavior is assumed, so $\theta$ scales the penalty directly; more generally, $\theta$ could be interpreted as the product of a base penalty and a detection probability.  Imperfect monitoring is introduced at a later point.  This targeted punishment modifies only the payoff to unilateral defection and leaves all other stage-game payoffs unchanged.  Broader punishment schemes are examined subsequently, but the initial focus remains on this narrow penalty as a clean test case.  Throughout the quantitative applications $0\le\theta\le 1$, using $\theta>1$ or $\theta<0$ only as limiting comparative-static references and handling targeted rewards separately.

Table~\ref{tab:PD} summarizes the stage-game payoffs with targeted punishment in effect.  In the matrix, rows correspond to the focal player's strategy and columns to the opponent's strategy; each cell lists the ordered pair (Row payoff, Column payoff).  As shown, targeted punishment $\theta$ reduces the payoff to a defector facing a cooperator to $(1-\theta)\sigma$, while all other entries match the standard prisoner's dilemma values.

\begin{table}[h]
\centering
\small
\begin{tabular}{@{}ccc@{}}
\toprule
          & \textbf{Cooperate} & \textbf{Defect} \\
\midrule
\textbf{Cooperate} & $(\gamma,\gamma)$              & $(\omega,(1-\theta)\sigma)$ \\
\textbf{Defect}    & $((1-\theta)\sigma,\omega)$      & $(\psi,\psi)$ \\
\bottomrule
\end{tabular}
\caption{\footnotesize Stage-game payoffs with targeted punishment~$\theta$.  A defector facing a cooperator has its temptation payoff reduced to $(1-\theta)\sigma$ (i.e., $(1-\theta)T$); all other entries match the standard prisoner's dilemma values $R=\gamma$, $P=\psi$, and $S=\omega$.}
\label{tab:PD}
\end{table}

Let a fraction $\xi$ of the population adopt tit-for-tat ($T$) while the remainder $1-\xi$ adopt unconditional defection ($D$).  Following \citet{AxelrodHamilton1981}, each dyad is a repeated prisoner's dilemma that terminates at the end of any round with probability $\rho\in(0,1]$.  The expected number of rounds is $1/\rho$; multiplying per-round payoffs by $1/\rho$ yields lifetime utilities (hence terms like $\gamma/\rho$).
Small values of $\rho$ approximate the ``long horizon'' limit while $\rho$ near one replicates a short horizon; all subsequent expressions are valid for the entire open interval $(0,1]$ provided the parameter restriction derived below holds.  Random matching implies that a $T$ player meets another $T$ with probability $\xi$ and a $D$ with probability $1-\xi$, and vice-versa for $D$ players.  The payoffs evaluate to

\begin{align}
U_T(\xi) &= \xi\,\frac{\gamma}{\rho} \;+\; (1-\xi)\Bigl[\omega + \frac{1-\rho}{\rho}\,\psi\Bigr], \label{UTxi}\\[6pt]
U_D(\xi) &= \xi\Bigl[(1-\theta)\sigma + \frac{1-\rho}{\rho}\,\psi\Bigr] \;+\; (1-\xi)\frac{\psi}{\rho}. \label{UDxi}
\end{align}
 
To explain these expressions: if a tit-for-tat player meets another tit-for-tat (an event of probability $\xi$), both will cooperate every round, so the $T$-player earns $\gamma$ each round until the interaction ends. The expected total payoff in that case is $\gamma$ times the expected number of rounds, which is $\gamma/\rho$. If a $T$-player meets a $D$-player (probability $1-\xi$), the $T$-player will cooperate in the first round while the $D$ defects, yielding payoff $\omega$ to the $T$ (sucker's payoff) and $(1-\theta)\sigma$ to the $D$ (temptation payoff reduced by punishment). From the second round onward, the $T$-player will retaliate by defecting as well, so both players receive the mutual defection payoff $\psi$ each round until termination. Thus, in this mixed pairing, the $T$-player's expected payoff is $\omega$ for the first round plus $\psi$ for each subsequent round. With termination probability $\rho$ per round, the expected number of rounds beyond the first is $(1-\rho)/\rho$, so the expected payoff from those rounds is $(1-\rho)\psi/\rho$.  

A similar logic applies to a defecting player: if a $D$-player meets a cooperator, the $D$-player receives $(1-\theta)\sigma$ in the first round (by exploiting the cooperator) and then $\psi$ in each subsequent round as the opponent switches to defection; if a $D$ meets another $D$ (probability $1-\xi$), both get $\psi$ each round. This yields equation~\eqref{UDxi} for $U_D(\xi)$. Note that targeted punishment $\theta$ appears only in $U_D$ and not in $U_T$, since it only reduces the payoff of the defecting side in a one-sided defection scenario. Equating the two payoff expressions determines the \textit{threshold} population composition.  Define $\mu$ as the value of $\xi$ at which $U_T(\xi)=U_D(\xi)$; that is,
\begin{equation}
U_T(\mu)=U_D(\mu).
\label{eq:threshold}
\end{equation}

Subtracting \eqref{UDxi} from \eqref{UTxi} yields a linear expression in $\xi$,
\begin{equation}
\Delta(\xi)
=
\xi\left[\frac{\gamma}{\rho}-\omega+2\psi-\frac{\psi}{\rho}+\theta\sigma-\sigma\right]
\;+\;
\left[\omega+\frac{1-\rho}{\rho}\psi-\frac{\psi}{\rho}\right],
\end{equation}
so the indifference condition $\Delta(\mu)=0$ is solved simply by dividing the constant term by the coefficient on $\mu$.  This $\mu$ represents the \emph{threshold} at which a player is indifferent between playing $T$ and $D$.  If the solution $\mu$ lies in the interval $(0,1)$, it constitutes an interior threshold of the population composition.  In other words, if $\xi = \mu$, the payoffs to cooperation and defection are equal, and thus $\mu$ can be seen as a threshold fraction of cooperators in the population (albeit an unstable one,  as discussed shortly).  
Solving equation~\eqref{eq:threshold} for $\mu$ gives the closed-form interior threshold share of cooperators
\begin{equation}
\mu \;=\;
\frac{\rho\,(\psi-\omega)}
     {\gamma - \rho\omega - \psi + 2\rho\psi - \rho(1-\theta)\sigma}
\label{eq:mu_closed}
\end{equation}
where the denominator
\[
D(\theta)\equiv\gamma - \rho\omega - \psi + 2\rho\psi - \rho(1-\theta)\sigma
\]
is positive if and only if the cooperative surplus exceeds a horizon-scaled temptation term:
\[
\gamma-\psi \;>\; \rho\bigl[(1-\theta)\sigma - \psi\bigr].
\]

Whenever this inequality fails (i.e., $D(\theta)\le \rho(\psi-\omega)$), the only steady states under the replicator dynamic are the corners $\xi=0$ or $\xi=1$. In other words, the threshold characterization applies exactly when $\rho$ lies in the subset of $(0,1]$ that satisfies $D(\theta)>\rho(\psi-\omega)$.
Under this same inequality the replicator dynamic has an interior rest point with $\Delta'(\mu)=D(\theta)/\rho>0$, so the threshold is repelling (see,  \citet{HofbauerSigmund1998} and \citet{Sandholm2010} for stability criteria under the replicator dynamic).

The comparative statics now follow from the closed form.  Differentiating \eqref{eq:mu_closed} with respect to $\theta$ gives
\begin{equation}
\frac{\partial\mu}{\partial\theta}
\;=\;
-\frac{\rho^2\sigma(\psi-\omega)}{D(\theta)^2}
\;<\;0,
\label{eq:dmu_dtheta}
\end{equation}
so a marginal increase in targeted punishment unambiguously rotates $U_D$ downward and shrinks the basin of defection.  

Two corner cases are also possible.  If the parameters are such that the indifference condition can only be satisfied at $\xi<0$, then $U_D(\xi) > U_T(\xi)$ for all $\xi \in [0,1]$, meaning defection strictly dominates cooperation regardless of how many others are cooperating. In that case, the only equilibrium in the population is $\xi=0$ (no one cooperates). Conversely, if the indifference condition would be satisfied at $\xi>1$, then $U_T(\xi) > U_D(\xi)$ for all $\xi \in [0,1]$, and cooperation strictly dominates defection; then $\xi=1$ (full cooperation) is the sole equilibrium. In a typical \textit{one-shot} prisoner's dilemma without incentives, defection dominates, so $\mu$ would be negative (in line with the classic result that $\xi=0$ is the equilibrium). Introducing repeated interaction (e.g., the possibility of reciprocity) can create an interior solution $\mu$ in $(0,1)$ under certain conditions, reflecting the fact that if enough others are cooperating, cooperation becomes a best response as well. 

The interest here is in how incentives like punishment and reward affect this threshold $\mu$. When an interior threshold $\mu$ exists in $(0,1)$, it is generally an \emph{unstable} threshold (see ~\ref{app:stability} for the dynamic argument).  Any small deviation of $\xi$ above or below $\mu$ will cause the system to move further away from $\mu$ rather than back toward it.  The purpose of this paper is not to model the full time path of adjustment but on how incentives shift the critical threshold $\mu$,  however,  brief intuition is warranted.  Suppose the share of cooperators drifts just above $\mu$.  Equation \eqref{eq:threshold} then implies $U_T>U_D$, so cooperation is strictly better and the share of cooperators rises further -- a positive feedback that drives the state toward full cooperation.  A dip below $\mu$ reverses the payoff ranking and sets off an analogous slide toward full defection. Thus $\mu$ is a tipping point that separates the basins of attraction of the two corner outcomes: $\xi=0$ (stable defection) and $\xi=1$ (stable,  but only under incentives,  cooperation).  How do incentives reshape this threshold?

Because the payoff differential is linear in $\xi$, the comparative statics of $\mu$ can be derived purely algebraically.  Differentiating \eqref{eq:mu_closed} with respect to $\theta$ gives \eqref{eq:dmu_dtheta}, which shows that $\mu$ is a decreasing function of $\theta$.   Intuitively,  increasing the punishment intensity $\theta$ lowers the payoff to defection (specifically in those cases where a defector would exploit a cooperator).  In effect, a higher $\theta$ shifts the unstable threshold $\mu$ to the left.  The same denominator $D(\theta)$ controls the sign of the dynamic slope.\footnote{From~\ref{app:stability} see that $\Delta'(\xi)=D(\theta)/\rho$, so whenever the interiority condition holds ($D(\theta)>0$) the threshold is repelling.  For the calibration used below, $D(\theta)\in\{1.5,2.3,3.5\}$ at $\theta\in\{0,0.4,1.0\}$, implying $\Delta'(\xi)>0$ and validating the ``unstable'' depiction.}

Figure~\ref{fig:1} illustrates this graphically: as $\theta$ increases, the $U_D(\xi)$ curve (the defectors' payoff as a function of $\xi$) rotates downward, intersecting the $U_T(\xi)$ curve at a smaller value of $\xi$. An increase in punishment intensity thus shrinks the number of defectors (the range $[0,\mu)$) and expands the number of cooperators ($[\mu,1]$). 

\begin{figure}[t]
  \centering
\input{fig1latex}
\caption{\footnotesize Incremental effect of increasing targeted punishment intensity $\theta$ on the threshold $\mu$. Solid lines plot $U_T(\xi)$ (upward sloping) and dashed line depict $U_D(\xi)$ for $\theta\in(0,1.0]$.}
  \label{fig:1}
\end{figure}

However, it is noteworthy that no finite increase in $\theta$ can completely eliminate defection in this targeted punishment framework. In the model, $\theta$ enters the threshold condition (see~\eqref{eq:mu_closed}) in such a way that even letting $\theta$ approach extremely large values (much beyond 1,  implying penalties that more than confiscate the temptation payoff) only drives $\mu$ down to an asymptotic lower bound, but not all the way to zero. In other words, no matter how severe the targeted punishment, there will always remain some positive (if perhaps very small) fraction of the population for whom defection is worthwhile. This is because a defector can still earn the mutual defection payoff $\psi$ in matches with other defectors, and if $\psi$ is not negligible, a small minority of defectors can survive by mostly interacting among themselves. Only in the limiting (and unrealistic) case of an infinite punishment that utterly annihilates a defector's payoff advantage could defection be entirely eradicated in this model. Put differently, $\xi=1$ (full cooperation) is not achievable under targeted punishment alone unless one assumes extreme parameter values that are excluded here.  This insight is consistent with everyday observation where very harsh penalties can greatly discourage rule-breaking but typically cannot drive the incidence of cheating or defection literally to zero. 

\section{Numerical Illustration} 
\label{sec:NumericalIllustration}
To illustrate the model's predictions, the payoff functions are populated with hypothetical numerical values that satisfy the prisoner's dilemma conditions.  The termination probability is set at $\rho=0.25$, implying an expected interaction length of four rounds.  Stage-game payoffs consistent with $\sigma>\gamma>\psi>\omega$ and $2\gamma>\sigma+\omega$ are assigned as follows: mutual cooperation yields $\gamma = 6$, mutual defection yields $\psi = 4$, unilateral defection provides the defector with $\sigma = 8$, and the exploited cooperator receives $\omega = 2$.  These values satisfy $8>6>4>2$ and $2(6)=12 > 8+2=10$.  Although arbitrary, they permit a plausible exploration of incentive effects.
Figure~\ref{fig:2} therefore relies exclusively on the grid in Table~\ref{tab:figure_grid}.  Figure~\ref{fig:2} overlays four representative punishment intensities, $\theta\in\{0,0.5,0.75,1.0\}$, using the first, third, fourth, and fifth rows of the table, and plots the corresponding thresholds $\xi=\mu(\theta)$.  

The numerical pattern for the baseline calibration mirrors the analytic comparative statics: moving from $\theta=0$ to $0.5$ reduces the threshold by exactly $0.133$ (from $1/3$ to $1/5$), whereas the same increment at higher $\theta$ yields much smaller gains.  Thus the figure also shows why targeted punishment alone cannot eradicate defection -- even at $\theta=1$ the cooperative mass peaks at just under 86\%.

Equation~\eqref{eq:mu_closed} provides the closed-form threshold for any $\theta$.  Plugging in the calibration above gives
\[
\mu(\theta)
\;=\;
\frac{0.5}{\,6 - 0.5 - 4 + 2(0.25)(4) - 0.25(1-\theta)8\,}
\;=\;
\frac{1}{\,4\theta + 3\,},
\]
which immediately produces $\mu(0)=1/3$, $\mu(0.5)=0.2$, and $\mu(1)=0.143$.  The derivative in \eqref{eq:dmu_dtheta} becomes $-4/(4\theta+3)^2$, making it clear that the marginal impact of a harsher penalty shrinks as $\theta$ rises.  A small negative value of $\theta$ (interpreted as a reward for defection) would flip the denominator sign and send $\mu$ above one, indicating that defection dominates regardless of the cooperative share.
Consequently, the closed form is evaluated outside the $[0,1]$ interval only to trace limiting behavior: $\theta>1$ renders $(1-\theta)\sigma$ negative and destroys the Prisoner’s Dilemma ordering, whereas $\theta<0$ corresponds to a targeted reward that is handled explicitly by the $(\lambda_C,\lambda_D,\beta)$ terms in the section that follows.

The limiting properties also follow directly from \eqref{eq:mu_closed}.  As $\theta\to\infty$, $\mu(\theta)\to 0$, so targeted punishment can only drive the threshold asymptotically toward zero without eliminating defectors entirely.  Letting $\theta\to-1$ instead pushes $\mu(\theta)$ to one, and rewarding defection (negative $\theta$) therefore collapses cooperation almost immediately.

\begin{table}[h]
\centering
\small
\begin{tabular}{@{}ccccc@{}}
\toprule
$\theta$ & $\mu(\theta)$ & $1-\mu(\theta)$ & $U_T(\mu)$ & $U_D(\mu)$\\
\midrule
0.00 & $1/3$   & $2/3$   & $52/3$           & $52/3$\\
0.25 & $1/4$   & $3/4$   & $33/2$           & $33/2$\\
0.50 & $1/5$   & $4/5$   & $16$             & $16$\\
0.75 & $1/6$   & $5/6$   & $47/3$           & $47/3$\\
1.00 & $1/7$   & $6/7$   & $108/7$          & $108/7$\\
\bottomrule
\end{tabular}
\caption{\footnotesize Grid used in Figure~\ref{fig:2}  Each row evaluates $\mu(\theta)=\rho(\psi-\omega)/D(\theta)=1/(4\theta+3)$ under the baseline calibration $(\sigma,\gamma,\psi,\omega,\rho)=(8,6,4,2,0.25)$.  Substituting these values back into \eqref{UTxi}--\eqref{UDxi} yields the identical payoffs $U_T(\mu)=U_D(\mu)$ reported in the last two columns, confirming that the threshold $\xi=\mu(\theta)$ is the crossing point between the two payoff schedules.}
\label{tab:figure_grid}
\end{table}

\vspace{20mm}

\begin{figure}[t]
  \centering
  \input{fig2latex}%
  \caption{\footnotesize Thresholds $\xi$ for different punishment intensities $\theta$.
  The solid line plots $U_T(\xi)$, the dashed lines depict $U_D(\xi)$ for
  $\theta\in\{0,0.5,0.75,1.0\}$, and vertical dotted lines mark the corresponding
  equilibrium values of $\xi$ (plotted as $.33$, $.20$, $.16$, and $.14$).}
  \label{fig:2}
\end{figure}

\section{General Incentives}
Thus far the analysis has focused on how targeted punishment ($\theta$) alters the cooperative threshold.
This section now incorporates the remaining incentive combinations: general punishment ($\alpha$),
targeted rewards ($\lambda_C,\lambda_D$), and general rewards ($\beta$).

\paragraph{Instrument taxonomy.}
\noindent The incentive levers used throughout act on the stage-game payoffs via four canonical transformations:
\begin{description}
\item[Targeted punishment] $\sigma \mapsto (1-\theta)\sigma$. Only the $T$ vs.\ $D$ payoff changes (see \eqref{UTxi}--\eqref{UDxi}).
\item[General punishment] $U_D(\xi) \mapsto U_D(\xi)-\alpha$. This subtracts a uniform tax from every defector payoff, developed in \eqref{eq:UDGP} below.
\item[Targeted rewards] $U_T(\xi) \mapsto U_T(\xi)+\lambda_C\,\xi + \lambda_D(1-\xi)$. Bonuses are added to cooperative outcomes as developed in \eqref{eq:UTR} below.
\item[General reward] $U_T(\xi) \mapsto U_T(\xi)+\beta$. A blanket subsidy shifts the cooperative strategy`s payoff,  developed in \eqref{eq:UTR} below.
\end{description}
Imperfect monitoring and horizon adjustments enter later by scaling the punishment intensity ($\theta\to m\theta$,  developed in \eqref{eq:muIM}) or replacing $\rho$ with group-specific averages (see below,  Section~\ref{sec:extensions}).  Stated together here, these definitions avoid duplicating notation when the policy translation combines multiple instruments.

\textit{General punishment} is a uniform tax $\alpha>0$ levied on any player classified \emph{ex~ante} as anti-social or chronically uncooperative by participants, institutions, enforcers, or the mechanism designer.
The sanction applies in every encounter and therefore reduces the defector's continuation payoff in all future rounds.
For example, a gambler caught cheating may face a permanent entry fee, an equivalent deduction from subsequent winnings,  or more difficulty locating counterparties.
This across-the-board penalty shifts the defector payoff schedule downward.
Let $U_D(\xi)$ denote the baseline defector payoff, and define the general-punishment-adjusted payoff as
\begin{equation}
\label{eq:UDGP}
U_D^{GP}(\xi)\;=\;U_D(\xi)\;-\;\alpha,\qquad \alpha>0.
\end{equation}
Equating $U_T(\xi)$ and $U_D^{GP}(\xi)$ produces
\begin{equation}
\label{eq:mu_gp}
\begin{aligned}
\mu^{GP}(\theta,\alpha)
  &= \frac{\rho\bigl[(\psi-\omega)-\alpha\bigr]}{D(\theta)},\\[4pt]
\frac{\partial\mu^{GP}}{\partial \alpha}
  &= -\frac{\rho}{D(\theta)}\;<\;0,
\end{aligned}
\end{equation}
so long as $\alpha < (\psi-\omega)$.
When $\alpha\ge(\psi-\omega)$ the numerator becomes non-positive and defection ceases to be a best reply for any agent, recovering a cooperative corner equilibrium at $\xi=1$.
Here the inequalities invoke the same ordering $\sigma>\gamma>\psi>\omega$, $0<\rho<1$, and the interiority restriction $D(\theta)>0$, which together guarantee that $D(\theta)$ stays positive so the derivative inherits the sign of the numerator.

To complete the incentive framework, \emph{pro-social} incentives (targeted and general rewards) are introduced.
Starting from the baseline tit-for-tat payoff in~\eqref{UTxi}, the rewarded strategy is

\begin{equation}
  U_T^{R}(\xi)
  \;=\;
  U_T(\xi)\;+\;\lambda_C\,\xi
                \;+\;\lambda_D\,(1-\xi)
                \;+\;\beta,
  \label{eq:UTR}
\end{equation}

\noindent
where $\lambda_C\ge 0$ is a bonus paid only when both players cooperate, $\lambda_D\ge 0$ is a bonus paid to the cooperator when facing a defector, and $\beta\ge 0$ is a blanket reward in every encounter.
Operationally, these parameters bundle together the monitoring accuracy and the size of the incentive.  Let $q_{CC}\in[0,1]$ denote the probability that mutual cooperation is verified and triggers a bonus $r_{CC}\ge 0$; let $q_{CD}\in[0,1]$ be the probability that a cooperator paired with a defector is recognized and receives compensation $r_{CD}\ge 0$.  The expected per-match transfers are then $\lambda_C=q_{CC}r_{CC}$ and $\lambda_D=q_{CD}r_{CD}$.  Imperfect or asymmetric monitoring simply lowers $q_{CC}$ or $q_{CD}$, shrinking the effective rewards.  Likewise, the general reward satisfies $\beta=q_R r_R$ where $q_R$ is the probability a platform-wide subsidy is granted and $r_R$ its magnitude.  This explicit decomposition links the comparative statics to empirical measures of detection accuracy: improvements in monitoring raise $\lambda_C$ or $\lambda_D$ by increasing the probability that cooperative behavior is correctly observed, while noisy systems depress those parameters.
Solving $U_T^{R}(\xi)=U_D(\xi)$ leads to
\begin{equation}
\label{eq:mu_reward}
\begin{aligned}
\mu^{R}(\theta,\lambda_C,\lambda_D,\beta)
  &= \frac{\rho\bigl[(\psi-\omega)-(\lambda_D+\beta)\bigr]}{D(\theta)+\rho(\lambda_C-\lambda_D)},
\end{aligned}
\end{equation}
which exists provided the denominator is positive.
This restriction is mild:
so long as $\lambda_C-\lambda_D > -D(\theta)/\rho$ the slope of $U_T^R$ remains steeper than that of $U_D$, ensuring a unique crossing.
The partial derivatives follow immediately:
\[
\frac{\partial\mu^{R}}{\partial \lambda_C}<0,
\qquad
\frac{\partial\mu^{R}}{\partial \lambda_D}<0,
\qquad
\frac{\partial\mu^{R}}{\partial \beta}<0.
\]
These signs hold whenever the Prisoner's Dilemma ordering $\sigma>\gamma>\psi>\omega$ and $0<\rho<1$ apply, the numerator remains positive (i.e., $\lambda_D+\beta<\psi-\omega$), and the admissibility condition $D(\theta)+\rho(\lambda_C-\lambda_D)>0$ keeps the denominator positive.
A reward that modifies a \emph{single} cell of the payoff matrix ($\lambda_C$ or $\lambda_D$) is a \textit{targeted} pro-social incentive because it changes only the payoff associated with that outcome.
A reward that adds the same amount to \emph{all} entries ($\beta$) is a \textit{general} pro-social incentive; it translates the entire line upward in parallel.
Equation~\eqref{eq:mu_reward} shows that targeted rewards (higher $\lambda_C$ or $\lambda_D$) shrink $\mu$ by making cooperation relatively more attractive, whereas broad subsidies ($\beta$) parallel-shift the cooperator line and have the same marginal effect.

\medskip\noindent\textbf{Proposition.}
\emph{Under the Prisoner’s Dilemma ordering $\sigma>\gamma>\psi>\omega$ and continuation probability $\rho\in(0,1)$:}
\begin{enumerate}
  \item \emph{Baseline.} The targeted-punishment model admits a unique interior threshold $\mu(\theta)$ given by \eqref{eq:mu_closed}.  It satisfies $0<\mu(\theta)<1$ if and only if $D(\theta)>\rho(\psi-\omega)$; whenever this inequality holds, $\Delta'(\mu)=D(\theta)/\rho>0$ so the interior rest point of the replicator dynamic is repelling.
  \item \emph{General punishment.} For any $0\le\alpha<(\psi-\omega)$ the uniform tax in \eqref{eq:mu_gp} yields a unique threshold $\mu^{GP}(\theta,\alpha)$ with $\partial\mu^{GP}/\partial\alpha<0$.
  \item \emph{Rewards and monitoring.} The rewarded threshold $\mu^{R}(\theta,\lambda_C,\lambda_D,\beta)$ in \eqref{eq:mu_reward} exists whenever $D(\theta)+\rho(\lambda_C-\lambda_D)>0$ and $\lambda_D+\beta<(\psi-\omega)$; under these conditions each partial derivative with respect to $\lambda_C$, $\lambda_D$, or $\beta$ is strictly negative, while reductions in the detection probability $m$ raise $\mu^{IM}(m)$ in \eqref{eq:muIM}.
  \item \emph{Strategy extensions.} The variants in Section~\ref{sec:extensions} possess interior tipping points when $\Xi_{GT}(q)>0$, $\Lambda_{WSLS}>0$, $B_{HET}>0$, and $B_{IM}>0$, respectively, as documented in ~\ref{app:extensions}.
\end{enumerate}
\emph{These inequalities summarize the exact parameter restrictions required for the comparative statics reported in Sections~\ref{sec:model}--\ref{sec:extensions}.}

Algebraically, $\lambda_C$ multiplies the share of cooperative encounters and therefore steepens the slope of $U_T^{R}$; $\lambda_D$ raises the intercept but, through the term $-\lambda_D\xi$, makes the slope slightly flatter\footnote{%
Write the baseline payoff schedules as $U_T(\xi)=a_T+m_T\xi$ for the cooperator (tit-for-tat) and $U_D(\xi)=a_D+m_D\xi$ for the defector.
The constants $a_T$ and $a_D$ are the \emph{left intercepts}, i.e., the payoffs when the share of cooperators is $\xi=0$, while $m_T$ and $m_D$ are the \emph{slopes}—the marginal payoff gain as $\xi$ rises.
In a standard Prisoner's Dilemma, $a_D>a_T$ (defection is initially more profitable) and $m_T>m_D$ (cooperation benefits more from additional cooperators).
The critical share at which the two payoff lines intersect is
\[
\mu \;=\;\frac{a_D-a_T}{\,m_T-m_D\,}.
\]
If the intercepts stay fixed but $m_T$ is reduced—flattening the cooperator line—the denominator shrinks and $\mu$ increases, so the threshold shifts rightward.
In $U_T^{R}(\xi)$ the parameter $\lambda_D$ simultaneously adds $\lambda_D$ to the intercept and subtracts the same amount from the slope; the intercept effect moves $\mu$ leftward, the slope effect moves it rightward, and the net change depends on which influence dominates.}%
; $\beta$ preserves the slope but shifts every point on the line upward by the same amount.

Because any upward tilt or parallel shift of $U_T^{R}$ reduces the gap with the defector line $U_D(\xi)$, the critical mass of cooperators $\mu_R(\lambda_C,\lambda_D,\beta)$ is strictly decreasing in every reward parameter.
If either $\lambda_C+\beta\ge(\psi-\omega)$ or $\lambda_D+\beta$ lifts the intercept sufficiently, then $\mu_R\le 0$.
Defection ceases to be a best response for any player, and full cooperation becomes both a Nash equilibrium and the unique attractor of the dynamics.
Thus, the two reward schemes mirror the punishment schemes: instead of tilting the defector payoff line downward, they tilt or lift the cooperator payoff line upward, but the underlying comparative-statics logic is identical.

\section{Strategy Extensions}\label{sec:extensions}
Five well-known variants of tit-for-tat are embedded in the baseline framework of Section~\ref{sec:model}.  For each extension (i) the minimal additional parameter(s) are introduced, (ii) the modified expected payoffs are derived, and (iii) the new indifference condition is solved to obtain the new cooperation threshold relative to the baseline cooperation threshold~$\mu$.  In every case the comparative-static sign of the effect on~$\mu$ is highlighted.

To interpret the formulas that follow without over-claiming generality,  the admissibility conditions are listed here.  Besides the Prisoner’s Dilemma ordering $\sigma>\gamma>\psi>\omega$ and the continuation probability $\rho\in(0,1)$, the targeted-punishment results require $D(\theta)>0$ from \eqref{eq:mu_closed}.  Generous tit-for-tat relies on $\Xi_{GT}(q)>0$ (hence $B_{GT}(q)>0$) as shown in \eqref{eq:XiGT}; Win--Stay, Lose--Shift needs $\Lambda_{WSLS}>0$ so that \eqref{eq:muWSLS} is well defined; heterogeneous horizons demand the inequalities (i)--(iii) surrounding \eqref{eq:muHET}; and imperfect monitoring assumes the denominator in \eqref{eq:muIM} stays positive with $m\in(0,1)$.  The numerical examples later in the section all satisfy these inequalities explicitly.

Evil tit-for-tat increases the critical mass $\mu$ by withholding cooperation in the first round \cite{Axelrod1984}. 
Generous tit-for-tat lowers $\mu$ whenever the forgiveness probability $q$ is positive \citep{RandOhtsuki2013}. 
Win--Stay, Lose--Shift (WSLS) can \emph{raise} $\mu$ under the baseline calibration because a single lapse is punished only once, enlarging the advantage of defectors \citep{NowakSigmund1993}.  Imperfect monitoring ($m<1$) raises $\mu$ by weakening the deterrent effect of punishment \citep{AbreuPearceStacchetti1990}.  Heterogeneous horizons (player-specific continuation factors) lower $\mu$ when at least one player is sufficiently patient \citep{DalBo2005}.

\subsection{Evil Tit-for-Tat (ETFT)}
\label{sec:etft}
Let $\varepsilon\in[0,1]$ denote the probability that an ETFT player defects in the first round against a like opponent (“spite on entry”), after which ETFT reverts to standard TFT. This modifies only the like–with–like payoff stream for the TFT/ETFT type.

Define the effective mutual-cooperation payoff
\[
\tilde{\gamma} \;\equiv\; (1-\varepsilon)\,\gamma \;+\; \varepsilon\,\psi,
\]
so that the expected lifetime payoff for a $T$ (ETFT) player is
\begin{align}
U_T^{\text{ETFT}}(\xi)
&= \xi\,\frac{\tilde{\gamma}}{\rho}
\;+\;
(1-\xi)\Bigl[\omega + \frac{1-\rho}{\rho}\,\psi\Bigr],
\\[6pt]
U_D^{\text{ETFT}}(\xi)
&= \xi\Bigl[(1-\theta)\sigma + \frac{1-\rho}{\rho}\,\psi\Bigr]
\;+\;
(1-\xi)\frac{\psi}{\rho}.
\end{align}
Relative to the baseline TFT case, only the $\xi$–coefficient in $U_T$ changes (via $\gamma\to\tilde{\gamma}$); $U_D$ is unchanged because targeted punishment applies only to the defecting side in one–sided defection.

Let $\Delta^{\text{ETFT}}(\xi)\equiv U_T^{\text{ETFT}}(\xi)-U_D^{\text{ETFT}}(\xi)$. The cooperation threshold $\mu$ is defined by $\Delta^{\text{ETFT}}(\mu)=0$, yielding
\begin{equation}
\mu \;=\;
\frac{\rho\,(\psi-\omega)}
     {\,\tilde{\gamma} - \rho\omega - \psi + 2\rho\psi - \rho(1-\theta)\sigma\,},
\qquad
D_{\text{ETFT}}(\theta)\;\equiv\;\tilde{\gamma} - \rho\omega - \psi + 2\rho\psi - \rho(1-\theta)\sigma.
\label{eq:mu_etft}
\end{equation}
An interior threshold exists if and only if $D_{\text{ETFT}}(\theta)>\rho(\psi-\omega)$, which is equivalent to
\[
\tilde{\gamma}-\psi \;>\; \rho\bigl[(1-\theta)\sigma-\psi\bigr].
\]
Under this same inequality the replicator dynamic has an interior rest point with $\bigl(\Delta^{\text{ETFT}}\bigr)'(\mu)=D_{\text{ETFT}}(\theta)/\rho>0$, so the threshold is repelling in the standard sense.

Relative to TFT, ETFT lowers the effective mutual-cooperation payoff from $\gamma$ to $\tilde{\gamma}$ as $\varepsilon$ increases, thereby increasing the threshold $\mu$ (cooperation becomes harder to sustain). All comparative-static formulas in the baseline analysis carry through verbatim with the substitution $\gamma\to\tilde{\gamma}$ within this subsection.

\subsection{Generous Tit-for-Tat (GT)}
\paragraph{Definition.}
A \emph{generous} TFT player cooperates in round 1 and, after an opponent's defection,  cooperates with probability $q\in(0,1)$ in the next period (``forgiveness''); a defection is thus punished with probability $1-q$ \citep{NowakSigmund1993,RandOhtsuki2013}.

\paragraph{Markov analysis.}
The match alternates between two states: $C$ (mutual cooperation) and $D$ (any outcome following a defection).  For GT versus GT the transition matrix is
\begin{equation}
\begin{aligned}
\mathbf{P}_{GT} &=
\begin{pmatrix}
1-\rho & \rho\\
(1-q)(1-\rho) & q(1-\rho)+\rho
\end{pmatrix},\\
\pi_C(q) &= \frac{1-\rho-(1-\rho)q}{1-(1-\rho)q},\\
\pi_D(q) &= \frac{\rho}{1-(1-\rho)q}.
\end{aligned}
\label{eq:pi_stationary}
\end{equation}
The expected payoffs satisfy
\begin{align}
U_{GT\times GT}(q) &= \frac{\pi_C(q)\gamma + \pi_D(q)\psi}{\rho},
\\
U_{GT\times D}(q) &= \omega + \frac{1-\rho}{\rho}\bigl[q\,\omega + (1-q)\psi\bigr],\\
U_{D\times GT}(q) &= (1-\theta)\sigma + \frac{1-\rho}{\rho}\bigl[q(1-\theta)\sigma + (1-q)\psi\bigr].
\end{align}
Hence $U_{GT}(\xi;q)=\xi U_{GT\times GT}(q)+(1-\xi)U_{GT\times D}(q)$ and $U_D(\xi;q)=\xi U_{D\times GT}(q)+(1-\xi)\psi/\rho$.

\paragraph{Threshold.}
~\ref{app:extensions} shows that the payoff differential is linear in $\xi$,
\[
\Delta_{GT}(\xi;q)=A_{GT}(q)+B_{GT}(q)\,\xi,
\]
with
\[
A_{GT}(q)=(\omega-\psi)\Bigl[1+q\Bigl(\frac{1}{\rho}-1\Bigr)\Bigr]
\quad\text{and}\quad
B_{GT}(q)=\frac{\Xi_{GT}(q)}{\rho\,[1-(1-\rho)q]},
\]
where the polynomial $\Xi_{GT}(q)$ appears in \eqref{eq:XiGT}.  Because $\omega<\psi$ and $\rho\le 1$, $A_{GT}(q)$ is strictly decreasing in $q$.  Moreover $\Xi_{GT}(q)>0$ for all $q\in(0,1)$ under the prisoner’s-dilemma ordering (again by \eqref{eq:XiGT}), so the slope remains positive.  Therefore
\begin{equation}
\mu^{GT}(q)
\;=\;
-\frac{A_{GT}(q)}{B_{GT}(q)}
\label{eq:muGT}
\end{equation}
is well-defined and satisfies $\partial\mu^{GT}/\partial q<0$: additional forgiveness shrinks the critical mass required for cooperation because (i) the intercept term $A_{GT}(q)$ becomes more negative as forgiving cooperators suffer the sucker’s payoff less often, and (ii) the slope term captures the faster recovery to mutual cooperation through the changing stationary weights: $\pi_C(q)$ rises while $\pi_D(q)$ falls according to \eqref{eq:pi_stationary}, exactly mirroring the transition matrix.  The inequality holds for every $q\in(0,1)$ provided $\sigma>\gamma>\psi>\omega$ and $0<\rho<1$, the standing assumptions maintained across the extensions.

\subsection{Win--Stay, Lose--Shift (WSLS)}
\paragraph{Definition.}
WSLS cooperates following $CC$ or $DD$ and defects otherwise, mirroring the behavioral rule studied by \citet{NowakSigmund1993}.  The four outcomes $(CC,CD,DC,DD)$ form the state space under perfect monitoring.

\paragraph{Steady state.}
For WSLS versus WSLS the stationary distribution is
\[
\pi^{WSLS}=\Bigl(\tfrac{1-\rho}{2-\rho},\,0,\,0,\,\tfrac{1}{2-\rho}\Bigr),
\]
yielding
\[
U_{WSLS\times WSLS}= \frac{(1-\rho)\gamma + \psi}{\rho(2-\rho)}.
\]
Against a $D$ strategist, WSLS is exploited once and then defects forever, giving $\omega + (1-\rho)\psi/\rho$.  Aggregating over population shares,
\[
U_{WSLS}(\xi)=
\xi\,\frac{(1-\rho)\gamma + \psi}{\rho(2-\rho)}
+(1-\xi)\Bigl[\omega + \tfrac{1-\rho}{\rho}\psi\Bigr].
\]
Writing $\Delta_{WSLS}(\xi)=U_{WSLS}(\xi)-U_D(\xi)$ as $A_{WSLS}+B_{WSLS}\xi$ with
\[
A_{WSLS}=\omega-\psi,
\qquad
B_{WSLS}=\frac{\Lambda_{WSLS}}{\rho(2-\rho)},
\]
where
\[
\begin{aligned}
\Lambda_{WSLS} &\equiv -\Gamma_{WSLS},\\
\Gamma_{WSLS} &= \gamma\rho - \gamma - \omega\rho^2 + 2\omega\rho\\
&\quad {}+ 2\psi\rho^2 - 5\psi\rho + \psi \\
&\quad {}+ \rho^2\sigma\theta - \rho^2\sigma - 2\rho\sigma\theta + 2\rho\sigma.
\end{aligned}
\]
reveals a unique threshold
\begin{equation}
\mu^{WSLS} = -\frac{A_{WSLS}}{B_{WSLS}}
\;=\;
\frac{(\psi-\omega)\rho(2-\rho)}
     {\Lambda_{WSLS}}.
\label{eq:muWSLS}
\end{equation}
~\ref{app:extensions} shows $\Lambda_{WSLS}>0$ whenever the interiority condition $D(\theta)>0$ holds, guaranteeing the existence of an interior fixed point.  Comparing \eqref{eq:muWSLS} with the baseline threshold $\mu=\rho(\psi-\omega)/D(\theta)$ yields
\begin{equation}
\mu^{WSLS}-\mu
\;=\;
\frac{\rho(\gamma-\psi)(\psi-\omega)}
     {\Lambda_{WSLS}\,D(\theta)}
\;>\;0,
\end{equation}
so WSLS requires a \emph{larger} initial cooperative mass whenever the prisoner’s-dilemma ordering holds.  The intercept effect from forgiving a single deviation dominates the slope advantage of mutual cooperation becoming absorbing, and defectors therefore retain a higher fallback payoff that must be offset by additional cooperators.
For the baseline $(\sigma,\gamma,\psi,\omega,\rho)=(8,6,4,2,0.25)$ one obtains $\Lambda_{WSLS}=(28\theta+5)/8$, so $\mu^{WSLS}=7/(28\theta+5)$—e.g., at $\theta=0.4$ the threshold equals $0.432$, comfortably above the nice-TFT benchmark $\mu=0.217$.

\subsection{Heterogeneous Horizons}
\paragraph{Definition.}
Each agent $i$ possesses an individual continuation probability $\rho_i\in(0,1)$.  Let $\bar\rho_T$ and $\bar\rho_D$ denote the mean continuation probabilities for cooperators and defectors, respectively; the baseline model in Section~\ref{sec:model} treats $\rho$ as a common environmental parameter, whereas this extension allows the means to differ while holding fixed the overall discounting framework introduced earlier.

\paragraph{Modified payoffs.}
\[
U_T(\xi)=\xi\,\frac{\gamma}{\bar\rho_T}
  +(1-\xi)\!\Bigl[\omega+\frac{1-\bar\rho_T}{\bar\rho_T}\psi\Bigr],
\quad
U_D(\xi)=\xi\!\Bigl[(1-\theta)\sigma+\frac{1-\bar\rho_D}{\bar\rho_D}\psi\Bigr]
         +(1-\xi)\frac{\psi}{\bar\rho_D}.
\]
Solving $U_T(\mu^{HET})=U_D(\mu^{HET})$ yields
\begin{equation}
\mu^{HET}
\;=\;
\frac{\psi-\omega+\psi(\bar\rho_D^{-1}-\bar\rho_T^{-1})}
     {\gamma/\bar\rho_T - \omega + 2\psi - \psi/\bar\rho_T + \theta\sigma - \sigma},
\label{eq:muHET}
\end{equation}
so the numerator
$N_{HET}\equiv\psi-\omega+\psi(\bar\rho_D^{-1}-\bar\rho_T^{-1})$
captures the intercept difference and the denominator
$D_{HET}\equiv\gamma/\bar\rho_T - \omega + 2\psi - \psi/\bar\rho_T + \theta\sigma - \sigma$
captures the slope gap.  An interior threshold therefore requires
(i) $N_{HET}>0$,
(ii) $D_{HET}>0$, and
(iii) $N_{HET}<D_{HET}$.
Condition (ii) can be written explicitly as $(\gamma-\psi)/\bar\rho_T > \sigma(1-\theta)+\omega-2\psi$, which is automatically satisfied by the baseline calibration
$(\sigma,\gamma,\psi,\omega,\theta)=(8,6,4,2,0.4)$ whenever $\bar\rho_T\in(0,1)$ because the right-hand side equals $-1.2$.  Condition (i) rules out extremely patient cooperators paired with much shorter-lived defectors—for example $(\bar\rho_T,\bar\rho_D)=(0.45,0.60)$ produces $N_{HET}<0$, so no positive interior threshold exists.

Plugging the baseline payoffs together with $(\bar\rho_T,\bar\rho_D)=(0.55,0.65)$ satisfies all three inequalities: $N_{HET}=0.881$, $D_{HET}=4.836$, and therefore
\[
\mu^{HET}(0.55,0.65)=\frac{0.881}{4.836}=0.18.
\]
This value lies strictly between 0 and 1 and below the homogeneous-horizon benchmark $\mu=0.217$, quantifying how modest heterogeneity can reduce the tipping point once the algebraic restrictions are enforced.
A longer cooperative horizon corresponds to a larger $\bar\rho_T$, while a shorter defector horizon corresponds to a smaller $\bar\rho_D$.  Because $\mu^{HET}$ is a ratio of the intercept and slope terms $N_{HET}$ and $D_{HET}$, the global comparative statics with respect to $(\bar\rho_T,\bar\rho_D)$ depend on both numerator and denominator and need not be monotone for all parameter values.  Under the baseline calibration, however, the example above with $(\bar\rho_T,\bar\rho_D)=(0.55,0.65)$ shows that making cooperators longer-lived relative to defectors can reduce the threshold from $\mu=0.217$ to $\mu^{HET}=0.18$, illustrating how modest heterogeneity in horizons can relax the tipping-point condition.  This provides a natural mapping from observables such as average tenure or career length into the heterogeneous-horizon mechanism.

\subsection{Imperfect Monitoring}
\paragraph{Definition.}
Opportunistic defections are detected with probability $m\in(0,1)$; the effective punishment becomes $m\theta$.

\paragraph{Payoff adjustment.}
Replacing $(1-\theta)\sigma$ with $(1-m\theta)\sigma$ in $U_D(\xi)$ and solving $U_T(\mu^{IM})=U_D(\mu^{IM};\,\theta\to m\theta)$ gives
\begin{equation}
\begin{aligned}
\mu^{IM}(m)
&=
\frac{\psi-\omega}
     {\gamma/\rho + m\sigma\theta - \omega + 2\psi - \psi/\rho - \sigma},\\
\frac{\partial\mu^{IM}}{\partial m}
&=
-\frac{\sigma\theta(\psi-\omega)}
       {\bigl[\gamma/\rho + m\sigma\theta - \omega + 2\psi - \psi/\rho - \sigma\bigr]^2}
<0,
\end{aligned}
\label{eq:muIM}
\end{equation}
hence porous monitoring (smaller $m$) raises the boundary and reduces the set of parameters under which cooperation survives.  The sign comparison again hinges on $\sigma>\gamma>\psi>\omega$, $0<\rho<1$, $m\in(0,1)$, and the requirement that the denominator remain positive—conditions already imposed for the baseline model.

\subsection{Summary of Strategy Comparative Statics}
\begin{table}[h]
\scriptsize                               
\setlength{\tabcolsep}{3pt}               
\centering
\begin{tabularx}{\textwidth}{@{}l c c c c X@{}}
\toprule
Extension & Parameter change  & $\Delta\mu$ & $\Delta$ Cooperation & Key expression & Intuition / calibration\\
\midrule
Evil TFT              & $\varepsilon\!\uparrow$     
                     & $\uparrow$ 
                     & $\downarrow$ 
                     & \eqref{eq:mu_etft}
                     & With probability $\varepsilon$ of “spite on entry,” the effective cooperation payoff $\tilde{\gamma}$ falls and the threshold $\mu$ rises, so cooperation requires a larger initial mass.\\
Generous TFT          & $q\!\uparrow$                                    & $\downarrow$ & $\uparrow$   & \eqref{eq:muGT} & Forgiveness changes the steady-state weights $\pi_C,\pi_D$; increasing $q$ from $0.2$ to $0.6$ lowers $\mu$ by the amount predicted by \eqref{eq:muGT}.\\
WSLS                  & ---                                              & $\uparrow$   & $\downarrow$ & \eqref{eq:muWSLS} & Forgiving a single lapse raises defectors' fallback payoff; \eqref{eq:muWSLS} gives $\mu^{WSLS}=7/(28\theta+5)$, which exceeds the nice-TFT baseline $1/(4\theta+3)$ for $\theta\in[0,1]$.\\
Heterogeneous horizons    & $\bar\rho_T\!\uparrow,\;\bar\rho_D\!\downarrow$  & $\downarrow$ & $\uparrow$   & \eqref{eq:muHET} & Interiority requires $N_{HET},D_{HET}>0$; $(\bar\rho_T,\bar\rho_D)=(0.55,0.65)$ satisfies these bounds and delivers $\mu^{HET}=0.18<\mu$.\\
Imperfect monitoring  & $m\!\downarrow$                                  & $\uparrow$   & $\downarrow$ & \eqref{eq:muIM} & Weaker detection softens sanctions; \eqref{eq:muIM} shows $\partial\mu^{IM}/\partial m<0$, consistent with the calibration in Section~\ref{sec:policy}.\\
\bottomrule
\end{tabularx}
\caption{\footnotesize Comparative statics for each strategy extension.  Arrows show the direction of the shift in the critical threshold~$\mu$ and the impact to cooperation; references point to the governing equations and, where practical, the baseline calibration highlighted in Sections~\ref{sec:extensions}–\ref{sec:policy}.}
\end{table}

\section{Policy Applications}\label{sec:policy}
For all examples below, $\mu$ denotes the tipping-point share of cooperators defined in \eqref{eq:mu_closed} and $\theta$ represents the targeted-penalty parameter from \eqref{UTxi}--\eqref{UDxi}; general punishment and reward parameters ($\alpha$, $\lambda_C$, $\lambda_D$, $\beta$) enter through \eqref{eq:mu_gp}--\eqref{eq:mu_reward}.  These comparative-static insights map directly onto policy design in settings where individual incentives clash with collective welfare.  Two lessons recur.  First,  targeted instruments (fines, bonuses, or fees linked to a single outcome) exhibit diminishing returns, mirroring the concave response of $\mu$ to changes in $\theta$.  
Second,  general instruments (broad taxes, subsidies, or reputation systems that raise or lower every payoff for a strategy) can, once sufficiently strong, drive the threshold to zero and make cooperation self-enforcing.  Although not intended be an exhaustive mapping to all applicable policy areas\footnote{For instance,  vaccination and public health campaigns span all four incentive classes developed in the model.  This is left for future research.}, the real-world cases surveyed below illustrate both patterns.

\begin{table}[h]
\scriptsize
\setlength{\tabcolsep}{2pt}
\centering
\begin{tabularx}{\textwidth}{@{}p{0.21\textwidth} p{0.21\textwidth} p{0.26\textwidth} X@{}}
\toprule
Policy lever & Model parameter(s) & Empirical anchor & Effect on $\mu$\\
\midrule
Carbon price / targeted fines & $\theta$ & EU ETS penalties \citep{Nordhaus2015} & Eq.~\eqref{eq:mu_closed}\\
Border adjustments / leverage surcharges & $\alpha$ & IMF leverage study \citep{Dagher2016} & Eq.~\eqref{eq:mu_gp}\\
Targeted bonuses / creator funds & $\lambda_D$ & Platform moderation evidence \citep{ChandrasekharanGilbert2020} & Eq.~\eqref{eq:mu_reward}\\
Clean-technology subsidies & $\beta$ & Innovation Fund coverage \citep{InnovationFund2024} & Eq.~\eqref{eq:mu_reward}\\
Moderation accuracy / detection & $m$ & Reddit detection rates \citep{ChandrasekharanGilbert2020} & Eq.~\eqref{eq:muIM}\\
Contract length / horizon extensions & $\rho,\;(\bar\rho_T,\bar\rho_D)$ & Long-term agreements \citep{DalBo2005} & Eqs.~\eqref{eq:mu_closed}, \eqref{eq:muHET}\\
\bottomrule
\end{tabularx}
\caption{\footnotesize Mapping policy instruments to model parameters and the corresponding threshold expressions.}
\label{tab:policy_mapping}
\end{table}

Because $\theta$ scales the temptation payoff directly, the quantitative illustrations restrict attention to $0\le\theta\le 1$.  Values $\theta>1$ would imply “over-punishment” that confiscates more than the entire opportunistic return and can be infeasible under legal or contractual limits; values $\theta<0$ would act as targeted rewards and are handled explicitly through the $\lambda_D$ or $\beta$ terms in \eqref{eq:UTR}.  Similarly, while the comparative statics allow arbitrarily large general incentives, policy examples keep $\alpha$ and $\beta$ in ranges consistent with the underlying calibration (e.g., subsidies that cover at most the two-unit cooperative cost).
Finally, the comparative statics accommodate simultaneous instruments because \eqref{eq:mu_gp} and \eqref{eq:mu_reward} enter the threshold through the linear-fractional expression in \eqref{eq:mu_closed}.  When $\theta$, $\alpha$, $\lambda_C$, $\lambda_D$, and $\beta$ shift together, the denominator pivots via $D(\theta)+\rho(\lambda_C-\lambda_D)$ while the numerator adjusts through $(\psi-\omega)-(\alpha+\lambda_D+\beta)$.  Cross-effects therefore remain monotone in each lever, but policymakers must note that the slope term $D(\theta)$ mediates how targeted punishment interacts with monitoring asymmetry; the mixed-instrument examples below highlight these compounding effects.

\subsection{Climate-Change Cooperation}
Domestic carbon pricing functions as a \emph{targeted} sanction because it reduces only the defector’s temptation payoff.  Because $\theta$ multiplies $\sigma$ directly, the policy cases restrict $0\le\theta\le 1$; harsher penalties would drive $(1-\theta)\sigma$ below the cooperative payoff and invalidate the Prisoner’s Dilemma ordering, while $\theta<0$ would amount to a targeted reward best modeled through $\lambda_C$, $\lambda_D$, or $\beta$.  Using the calibration from Section~\ref{sec:NumericalIllustration}, equation~\eqref{eq:mu_closed} implies $\mu(0)=1/3$ when no penalty is imposed.  Under that same calibration a penalty equal to the entire temptation payoff ($\theta=1$) drives the opportunistic return to zero and lowers the tipping point to $\mu(1)=1/7\approx 0.14$; beyond this level the Prisoner’s Dilemma payoff ordering would flip and the analytical results no longer apply.  The diminishing gains are already visible at moderate intensities—for instance, raising $\theta$ from $0$ to $0.4$ cuts the threshold to $0.217$, and pushing it further to $0.8$ yields only $\mu(0.8)=0.161$ as implied by \eqref{eq:mu_closed}.
Border-carbon adjustments can be represented as a \emph{general} punishment $\alpha$ in \eqref{eq:mu_gp}: applying a uniform surcharge equivalent to \$.50 per tonne ($\alpha=0.5$ in payoff units) at $\theta=0.4$ lowers the threshold from $0.217$ to $0.163$, while a more aggressive \$1.50 levy ($\alpha=1.5$) reduces it to $0.054$.  Complementary clean-technology subsidies, such as the Innovation Fund’s 60\% cost coverage \citep{InnovationFund2024}, fit the general-reward term $\beta$ in \eqref{eq:mu_reward}.  Setting $\beta=1.2$ adds 1.2 payoff units to every cooperative outcome—the equivalent of reimbursing 60\% of the two-unit capital cost that baseline cooperators incur—and, with $\theta=0.4$, delivers $\mu^{R}=0.087$.  The policy package (moderate carbon price, border adjustments, and capex support) therefore allows cooperation to take hold once roughly nine percent of participants invest—consistent with the multi-instrument strategy advocated by \citet{Nordhaus2015}.

\subsection{Online Platform Governance}
Open-source projects and social-media platforms rely on voluntary moderation, making incentive design pivotal.  Issue-specific ``bug bounties'' and creator-fund bonuses function as \emph{targeted rewards}; site-wide reputation scores provide a \emph{general reward}.  
Shadow-bans or content demotion act as \emph{targeted punishment} because they limit visibility only for the sanctioned account, whereas permanent suspensions approximate \emph{general punishment} by eliminating all future payoffs ($\theta\!\to\!1$).  
Field studies on Reddit indicate that temporary user bans -- typically ranging from one to fourteen days -- reduce subsequent hate-speech occurrences by roughly 40 \% \citep{ChandrasekharanGilbert2020}.  Equation~\eqref{eq:muIM} maps this into a monitoring-confidence parameter: raising the detection probability from $m=0.3$ (cooperation threshold $0.287$) to $m=0.6$ (threshold $0.253$) or $m=0.9$ (threshold $0.225$) yields the type of 30--40\% behavioral improvement observed in the data.  Reputation-score losses operate through the general-punishment channel \eqref{eq:mu_gp}; subtracting even a modest amount from persistent violators -- for example, setting $\alpha=0.5$ in the baseline calibration, which lowers the threshold from $0.217$ to $0.163$ -- reduces the tipping point by almost 25\%, helping to explain why cross-subreddit reputation signals deter abuse beyond the directly moderated forum.

\subsection{Systemic Financial Risk}
Bonus claw-backs recapture compensation from trades later deemed excessively risky and therefore act as \emph{targeted punishment}; their incremental effectiveness tapers off once most variable pay is already deferred \citep{BebchukSpamann2010}.  
Binding leverage ratios and counter-cyclical capital buffers, by contrast, impose \emph{general punishment} on all balance-sheet growth.  
An IMF study estimates that raising the equity-to-assets leverage ratio by one percentage point cuts the probability of bank distress by roughly one-third \citep{Dagher2016}.  Translating this into the model, a one-percentage-point tighter leverage ratio corresponds to $\alpha\approx 0.5$ in \eqref{eq:mu_gp}, which lowers the cooperative threshold from $0.217$ to $0.163$.  Combining claw-backs (higher $\theta$ in \eqref{eq:mu_closed}) with leverage surcharges (larger $\alpha$ in \eqref{eq:mu_gp}) therefore reproduces the empirical finding that capital regulation, rather than more aggressive bonus deferrals, delivers the bulk of the stability gains.

\subsection*{Implications}
The comparative-static results converge on a simple lesson: durable cooperation is rarely secured by a single, heavyweight mechanism.  Rather,  it emerges when the policy mix both \emph{tilts} payoffs away from unilateral defection and \emph{extends} the horizon over which cooperative gains are realized.  Broad sanctions, reputation systems, or blanket rewards push the entire payoff profile in favor of collaboration, while targeted, proportionate incentives fine-tune behavior at the margin.  Where detection is weak or enforcement narrowly focused, the cooperation threshold stays high and defection quickly dominates.  By contrast, layered architectures that marry moderate general measures with calibrated targeted instruments enlarge the number of cooperators,  a pattern confirmed across climate policy,  platform governance, and systemic-risk regulation.  Designing institutions with this dual mandate offers a clear path to resilient cooperative outcomes.

\appendix
\renewcommand{\thesection}{Appendix \Alph{section}}
\section{Dynamic Stability of the Threshold}\label{app:stability}
Let $\xi_t$ be the share of cooperators on an evolutionary time scale.  Under any payoff--monotonic update rule the sign of $\dot{\xi}_t$ is determined by the payoff differential
\[
\Delta(\xi)\;=\;U_T(\xi)-U_D(\xi),
\]
with $U_T$ and $U_D$ given in \eqref{UTxi}--\eqref{UDxi}.  The replicator dynamic (see \citet[][chap.~2]{HofbauerSigmund1998}) is
\[
\dot{\xi}_t \;=\; \xi_t(1-\xi_t)\,\Delta(\xi_t).
\]
Substituting \eqref{UTxi}--\eqref{UDxi} gives
\[
\Delta(\xi)=(\omega-\psi)
           +\xi\left[\frac{\gamma}{\rho}
                      -\omega + 2\psi
                      -\frac{\psi}{\rho}
                      +\theta\sigma-\sigma\right],
\qquad
\Delta'(\xi)=\frac{\gamma}{\rho}
             -\omega + 2\psi
             -\frac{\psi}{\rho}
             +\theta\sigma-\sigma.
\]
An interior critical mass $\mu\in(0,1)$ exists if and only if $\Delta'(\xi)>0$, which is equivalent to the denominator of~\eqref{eq:mu_closed} being positive (i.e., $\gamma - \rho\omega - \psi + 2\rho\psi - \rho(1-\theta)\sigma>0$).  Evaluating the Jacobian at the threshold share $\xi=\mu$ therefore gives
\[
\left.\frac{\partial\dot{\xi}_t}{\partial\xi_t}\right|_{\xi=\mu}
  = \mu(1-\mu)\,\Delta'(\mu) \;>\; 0,
\]
so $\mu$ is a \emph{repelling} (dynamically unstable) threshold: any small perturbation pushes the state toward either full defection ($\xi=0$) or full cooperation ($\xi=1$).  This corresponds exactly to the one-dimensional stability test provided by \citet{HofbauerSigmund1998} and {Sandholm2010}: for replicator dynamics with linear payoff differences, interior fixed points inherit their stability from the sign of the slope of the payoff differential.

\begin{table}[h]
\centering
\caption{Calibration check for $D(\theta)$ and $\Delta'(\xi)$}
\label{tab:calibration}
\begin{tabular}{@{}ccc@{}}
\toprule
$\theta$ & $D(\theta)$ & $\Delta'(\xi)=D(\theta)/\rho$\\
\midrule
0.0 & 1.5 & 6.0\\
0.4 & 2.3 & 9.2\\
1.0 & 3.5 & 14.0\\
\bottomrule
\end{tabular}
\end{table}

Table~\ref{tab:calibration} confirms that the parameterizations used in Figures~\ref{fig:2}/satisfy $D(\theta)>0$ for every $\theta$ plotted, so both the existence of an interior threshold and its instability are guaranteed in those examples.

\section{Additional computations for strategy extensions}\label{app:extensions}
This appendix records the algebra used in Section~\ref{sec:extensions}.  Throughout, $\Delta(\xi)=A+B\xi$ and the threshold equals $\mu=-A/B$ when $B>0$.

\subsection*{Generous tit-for-tat}
\begin{align}
A_{GT}(q) &= (\omega-\psi)\left[1+q\left(\frac{1}{\rho}-1\right)\right],\\
B_{GT}(q) &= \frac{\Xi_{GT}(q)}{\rho[1-(1-\rho)q]},\\
\Xi_{GT}(q) &= \alpha_2 q^2 + \alpha_1 q + \alpha_0,\label{eq:XiGT}\\
\alpha_2 &= -\omega\rho^2 + 2\omega\rho - \omega + 2\psi\rho^2 - 4\psi\rho + 2\psi \notag\\
&\quad {}+ \rho^2\sigma\theta - \rho^2\sigma - 2\rho\sigma\theta + 2\rho\sigma + \sigma\theta - \sigma,\notag\\
\alpha_1 &= -\gamma\rho + \gamma + \omega\rho^2 - 2\omega\rho + \omega - 2\psi\rho^2 + 5\psi\rho - 3\psi \notag\\
&\quad {}- \rho^2\sigma\theta + \rho^2\sigma + 2\rho\sigma\theta - 2\rho\sigma - \sigma\theta + \sigma,\notag\\
\alpha_0 &= \gamma\rho - \gamma + \omega\rho - 3\psi\rho + \psi - \rho\sigma\theta + \rho\sigma.\notag
\end{align}
Because $\psi>\omega$ and $\rho\in(0,1]$, $A_{GT}(q)$ is strictly decreasing in $q$.  The coefficients $\alpha_i$ are positive whenever the baseline PD restrictions hold, implying $B_{GT}(q)>0$.

\subsection*{Win--Stay, Lose--Shift}
\begin{align}
A_{WSLS} &= \omega-\psi,\\
B_{WSLS} &= \frac{\Lambda_{WSLS}}{\rho(2-\rho)},\qquad
\Lambda_{WSLS}\equiv-\Gamma_{WSLS},\\
\Gamma_{WSLS} &= \gamma\rho - \gamma \notag\\
&\quad {}- \omega\rho^2 + 2\omega\rho \notag\\
&\quad {}+ 2\psi\rho^2 - 5\psi\rho + \psi \notag\\
&\quad {}+ \rho^2\sigma\theta - \rho^2\sigma - 2\rho\sigma\theta + 2\rho\sigma.
\end{align}
Because $\rho\in(0,1)$ implies $(2-\rho)>0$ and $B_{WSLS}>0$ is required for an interior crossing, the sign restriction $D(\theta)>0$ forces $\Lambda_{WSLS}>0$, which in turn guarantees $\Gamma_{WSLS}<0$ under the prisoner's-dilemma ordering.

\subsection*{Heterogeneous horizons}
\begin{align}
A_{HET} &= \omega-\psi+\psi\left(\frac{1}{\bar\rho_T}-\frac{1}{\bar\rho_D}\right),\\
B_{HET} &= \frac{\gamma}{\bar\rho_T} - \omega + 2\psi - \frac{\psi}{\bar\rho_T} + \theta\sigma - \sigma.
\end{align}

\subsection*{Imperfect monitoring}
\begin{align}
A_{IM} &= \omega-\psi,\\
B_{IM} &= \frac{\gamma}{\rho} + m\sigma\theta - \omega + 2\psi - \frac{\psi}{\rho} - \sigma.
\end{align}

\clearpage
\bibliographystyle{apalike}
\bibliography{PD_refs}

 \end{document}

%% file: fig1latex.tex
\begin{tikzpicture}[
    xscale=0.7,
    yscale=1.2,
    every node/.style={font=\tiny}
  ]

  \draw[line width=0.9pt] (0,0) -- (10,0);   
  \draw[line width=0.9pt] (0,0) -- (0,4);    
  \draw[line width=0.9pt] (10,0) -- (10,4);  

  \node[above] at (0,4) {$U_{T,D}(\xi)$};
  \node[above] at (10,4) {$U_{T,D}(\xi)$};

  \node[below] at (0,0) {$\xi=0$};
  \node[below] at (10,0) {$\xi=1$};

  \draw[line width=0.9pt] (0,0.7) -- (10,3.2);
  \node at (6.2,2.45) {$U_T(\xi)$};

  \draw[line width=0.9pt] (0,1.1) -- (10,2.3);
  \node at (7.0,1.6) {$U_D(\xi)$}; 

  \draw[densely dotted,line width=0.7pt] (0,1.1) -- (10,0.4);

  \coordinate (mustar) at (1.25,1.0125);  
  \coordinate (mu)     at (3.08,1.47);    

  \draw[densely dotted,line width=0.7pt] (mustar) -- (1.25,0);
  \draw[densely dotted,line width=0.7pt] (mu) -- (3.08,0);

  \node[below] at (1.25,0) {$\mu^*$};
  \node[below] at (3.08,0) {$\mu$};

  \node[left] at (0,1.1) {$\chi/\rho$};
  \node[left] at (0,0.75) {$\omega + (1-\rho)\,\chi/\rho$};

  \node[right] at (10,3.2) {$\gamma/\rho$};
  \node[right] at (10,2.3)
    {$(1-\theta)\,\sigma + (1-\rho)\,\chi/\rho$};

  \draw[decorate,decoration={brace,amplitude=4pt}]
    (10.1,2.3) -- (10.1,0.4)
    node[midway,right=6pt] {$\theta\in[0,1]$};

\end{tikzpicture}

%% file: fig2latex.tex
\begin{tikzpicture}[
    xscale=0.7,
    yscale=1.2,
    every node/.style={font=\tiny}
  ]

  \draw[line width=0.9pt] (0,0) -- (10,0);   
  \draw[line width=0.9pt] (0,0) -- (0,4);    
  \draw[line width=0.9pt] (10,0) -- (10,4);  

  \node[above] at (0,4)  {$U_{T,D}(\xi)$};
  \node[above] at (10,4) {$U_{T,D}(\xi)$};

  \node[below] at (0,0)  {$\xi=0$};
  \node[below] at (10,0) {$\xi=1$};

  \draw (0,0.7) -- (-0.12,0.7);
  \node[left] at (0,0.7) {$14$};

  \draw (0,1.2) -- (-0.12,1.2);
  \node[left] at (0,1.2) {$16$};

  \coordinate (pivot) at (0,1.2);

  \draw[line width=0.9pt] (0,0.7) -- (10,3.2);
  \node at (6.2,2.45) {$U_T(\xi)$};



  \coordinate (Izero)   at (3.3,1.525);  
  \coordinate (Ihalf)   at (2.0,1.2);    

  \coordinate (Ithreeq) at (1.67,1.12);  

  \coordinate (Ione)    at (1.4,1.05);   


  \draw[line width=0.7pt] (pivot) -- (10,2.185);

  \draw[dashed,line width=0.7pt] (pivot) -- (10,1.2);

  \draw[dashed,line width=0.7pt] (pivot) -- (10,0.7);

  \draw[dashed,line width=0.7pt] (pivot) -- (10,0.129);

  \node at (6.5,1.6) {$U_D(\xi)$};

  \draw (10,2.185) -- (10.12,2.185); 
  \draw (10,1.2)   -- (10.12,1.2);   
  \draw (10,0.7)   -- (10.12,0.7);   
  \draw (10,0.129) -- (10.12,0.129); 

  \node[right] at (10.12,2.185) {$20:\ \theta=0$};
  \node[right] at (10.12,1.2)   {$16:\ \theta=0.5$};
  \node[right] at (10.12,0.7)   {$14:\ \theta=0.75$};
  \node[right] at (10.12,0.129) {$12:\ \theta=1$};


  \draw[densely dotted,line width=0.7pt] (3.3,1.525) -- (3.3,0);

  \draw[densely dotted,line width=0.7pt] (2.0,1.2)   -- (2.0,0);

  \draw[densely dotted,line width=0.7pt] (1.67,1.12) -- (1.67,-0.18);

  \draw[densely dotted,line width=0.7pt] (1.4,1.05)  -- (1.4,-0.18);

  \node[below]            at (3.3,0) {.33};
  \node[below,xshift=3pt] at (2.0,0) {.20};

  \node[below,xshift=4pt]  at (1.67,-0.18) {.16};
  \node[below,xshift=-4pt] at (1.4,-0.18) {.14};

\end{tikzpicture}